\documentclass[10pt]{iopart}
\usepackage{graphicx}
\usepackage{amssymb}

\newcommand{\bk}{{\bf k}}

\newcommand{\chib}{\overline{\chi}}

\newcommand{\phib}{\overline{\phi}}

\newcommand{\Mpl}{m_{\rm  Pl.}}

\newcommand{\vk}{{\bf k}}

\newcommand{\vx}{{\bf x}}

\newcommand{\vn}{{\bf n}}

\newcommand{\dd}{{\rm d}}

\newcommand{\mG}{{\cal G}}
\newcommand{\mB}{{\cal B}}

\newcommand{\mP}{{\cal P}}

\newcommand{\mS}{{\cal S}}
\newcommand{\mL}{{\cal L}}
\newcommand{\mM}{{\cal M}}

\newcommand{\dphi}{\varphi}
\newcommand{\vphi}{{\phi}}

\begin{document}

\title[Gravity and non-gravity mediated couplings in multiple-field inflation]{Gravity and non-gravity mediated couplings in multiple-field inflation}

\author{Francis Bernardeau}

\address{CEA, Institut de Physique Th{\'e}orique, 91191 Gif-sur-Yvette c{\'e}dex, France 
CNRS, URA-2306, 91191 Gif-sur-Yvette c{\'e}dex, France}
\ead{francis.bernardeau@cea.fr}
\begin{abstract}
Mechanisms for the generation of primordial non-Gaussian metric fluctuations in the context of multiple-field inflation are reviewed. 
As long as kinetic terms remain canonical, it appears that nonlinear couplings inducing non-gaussianities can be split into two 
types. The extension of the one-field results to multiple degrees of freedom leads to \textsl{gravity} mediated couplings that are ubiquitous but generally modest. Multiple-field inflation offers however the possibility of generating \textsl{non-gravity} mediated coupling in isocurvature directions that can eventually induce large non-Gaussianities in the metric fluctuations. The robustness of the predictions of such models is eventually examined in view of a case study derived from a high-energy physics construction.
\end{abstract}

\pacs{98.80.-k, 98.65.-r, 98.80.Bp, 98.80.Cq, 98.80.Es}

\section{Introduction}

It is now clearly understood that standard single field inflation cannot produce significant non-Gaussianities (NG)
during or immediately after the inflationary phase. The result obtained by Maldacena in Ref. \cite{2003JHEP...05..013M} explicitly shows that 
standard single field inflation leads to no or very little primordial non-Gaussianities. And this result appears to be very robust,  independent on the details of the model. This point is best illustrated by the expression of the bispectrum in the squeezed limit.
Defining the time dependent curvature modes $\zeta(t,\vk)$ and taking advantage of the statistical isotropy of the universe,
the power spectrum $\mP_{\zeta}$ of the field $\zeta$ can be defined as 
\begin{equation}
\langle\zeta(t,\vk_{1})\zeta(t,\vk_{2})\rangle=(2\pi)^3\delta_{\rm Dirac}(\vk_{1}+\vk_{2})\,\mP_{\zeta}(k_{1},t)
\end{equation}
and its bispectrum\footnote{In this context, both the power spectrum and the bispectrum will eventually be time independent at super-Hubble scales.}  $\mB_{\zeta}$ as 
\begin{equation}
\langle\zeta(t,\vk_{1})\zeta(t,\vk_{2})\zeta(t,\vk_{3})\rangle=(2\pi)^3\delta_{\rm Dirac}(\vk_{1}+\vk_{2}+\vk_{3})\,\mB_{\zeta}(\vk_{1},\vk_{2},\vk_{3},t).
\end{equation}
In the squeezed limit, i.e. when $k_{1}\ll k_{2}\approx k_{3}$, the bispectrum scales like $(n_{s}-1) \mP_{\zeta}(k_{1}) \mP_{\zeta}(k_{2})$ where $n_{s}$ is the spectral index \cite{2004JCAP...10..006C}. Not only are the nonlinear couplings naturally small -- say of order unity\footnote{Note however that the amount of NGs determined by a dimensionless quantity such as $\mB/\mP^{3/2}$ is of the order of $\mP_{*}(n_{s}-1)$ where $\mP_{*}\approx 10^{-5}$ is the amplitude of the metric fluctuations.} -- they are even suppressed by the slow-roll parameters (that ensures that $n_{s}$ is close to unity).

There are then two possible strategies to escape the limits set by Maldacena's results. One can modify the kinetic term by introducing higher order terms in the action that are not due to the potential shape. An example is provided by the  Dirac-Born-Infeld action \cite{2004PhRvD..70j3505S}. Such models will succeed in producing large NGs if precisely the kinetic term is, at the time of horizon crossing, at a non standard running point. That does not change however the squeezed limit case but allows large NG couplings for more equilateral type configurations of modes. This has been put forward as a powerful way for discriminating models \cite{2004JCAP...08..009B}. 

Another way of evading the constraints of standard single field inflation is to introduce multiple scalar degrees of freedom. It can actually be argued that this is a natural hypothesis since it is unlikely that only one fundamental degree of freedom will be light (e.g. compared to the Hubble energy scale) during the epoch of inflation. What is more hypothetical is whether those extra degrees of freedom can have observational consequences. By definition, degrees of freedom that do not participate in the metric fluctuation, at a given time, are called isocurvature modes. There is no reason why the isocurvature modes should remain so all along the history of the universe and various mechanisms have been put forward that can lead to a transfer of modes, from isocurvature to adiabatic modes.

For instance the curvaton model is based on the survival of (massive) isocurvature modes until late after the end of inflation that can alter the subsequent expansion history of the universe \cite{2002PhLB..524....5L}. This is a particular case of modulated inflation \cite{2003astro.ph..3614K,PhysRevD.69.023505,2004PhRvD..70h3004B}. Other mechanisms assume that isocurvature modes can change the end-point of inflation or alter the (p)-reheating effects (see \cite{2009PhRvL.103g1301B} and contribution by A. Frolov, this volume). Such mechanisms can also happen in the context of hybrid inflation. It does not mean yet that it induces non-Gaussian metric fluctuations. That would happen only if there are nonlinearities in the isocurvature-curvature transfer or if isocurvature modes are intrinsically NG at the time of transfer. This latter situation is in particular advocated in Refs. \cite{1990PhRvD..42.3936S,2002PhRvD..65j3505B,2002PhRvD..66j3506B,2003PhRvD..67l1301B} where isocurvature modes are shown to be able to develop large NG after horizon crossing. The aim of this paper is to show how different models that have been put forward  in the literature differ in their mechanisms for producing NGs and how they differ in the amplitude and/or shapes of NGs they produce. This will be described in section 2 where it is argued that one can distinguished between gravity and non-gravity mediated contributions.

The mere construction of working mechanisms cannot however be fully satisfactory. It is now clear that models can lead to 
a variety of observational signatures. However,  whether there exist natural realizations for those models from high-energy physics point of view is largely open. That will be tentatively addressed in section 3.

\section{From single to multiple-field inflation}

The class of models we are interested in corresponds to action that takes the form,
\begin{equation}
\mS=\int\dd\vx^3\dd t\sqrt{-g}\left[\frac{R\ \Mpl^2}{2}-\frac{1}{2}\partial_{\mu}\Phi_{a}\partial^{\mu}\Phi_{a}-V(\Phi_{1},\dots,\Phi_{N})\right]
\end{equation}
assuming that the $N$ scalar fields $\Phi_{a}$ are all minimally coupled to the metric and that they all have standard kinetic terms. Here, and in the following, summation over repeated latin indices $a, b, c, \dots$ from 1 to $N$ is implicitly assumed; $\Mpl$ is the reduced Planck mass, $\Mpl^{-2}=8\pi\,G$. We will also assume that the spatial sections are Euclidean.

As usual we will then assume that a semi classical approach can be used, e.g. that the fields $\Phi_{a}$ can be decomposed into an inhomogeneous part  $\vphi_{a}$ and a space dependent part $\dphi_{a}$ which, together with the scalar parts of the metric fluctuations, can be quantized (e.g. \cite{1992PhR...215..203M}). 

The zeroth order motion equation defines the field trajectory that is the time dependence of $\vphi_{a}$. When slow-rolling is reached it defines an adiabatic direction $\vn$ in the field space. More precisely we can define $\vn_{a}$ as a unit vector with,
\begin{equation}
\vn_{a}=\frac{\dot\vphi_{a}}{\sqrt{\dot\vphi_{b}^2}}\label{ndef}
\end{equation}
so that $\vn$ is tangential to the field trajectory during slow-roll.

The statistical properties of the metric field and/or perturbations will then be obtained from an expansion of the action around the homogeneous evolution,
\begin{equation}
\mS=\frac{1}{2}\int\dd\vx^3\dd t\ a^3\left[\mL_{0}+\mL_{2}+\mL_{3}+\dots
\right]
\end{equation}
where $\mL_{p}$ is of order $p$ in the field fluctuations\footnote{Up to a total derivative term, $\mL_{1}$ vanishes about the homogeneous trajectory.}.

\subsection{The equations}

From a practical point of view the choice of gauge can be crucial. Here we chose a spatially flat slicing gauge and derive the motion equation using the ADM formalism. This technique is very efficient and has been presented (and used) in Ref. \cite{2003JHEP...05..013M} in this context. We refer the reader to this paper for a detailed presentation of the method. 
In the ADM formalism the metric is written,
\begin{equation}
\dd s^2=-N^2\dd t^2+h_{ij}(\dd x^{i}+N^{i}\dd t)(\dd x^{j}+N^{j}\dd t),
\end{equation}
and we make the gauge choice so that
\begin{equation}
h_{ij}=a^2(t)\delta_{ij},
\end{equation}
introducing a spatially flat slicing gauge.

The resulting second order action is then given by,
\begin{eqnarray}
\mL_{2}&=&-V_{,ab}\,\dphi_{a}\dphi_{b}+\dot{\dphi}_{a}^2-\frac{1}{a^2}\left({\partial_{i}\dphi_{a}}\right)^2\nonumber\\
&&-\frac{2}{H{\Mpl^2}}\left(V_{,a}\dphi_{a}\right)(\dot\vphi_{a}\dphi_{a})
-\frac{V}{H^2{\Mpl^4}}\left(\dot\vphi_{a}\dphi_{a}\right)^2,
\label{action2nd}
\end{eqnarray}
The motion equation of the field $\varphi_{a}$ can then be read out of the action,
\begin{equation}
\ddot{\dphi}_{a}+3H\dot{\dphi}_{a}-\frac{1}{{a^2}}{\Delta}\dphi_{a}=s_{ab}\dphi_{b}
\end{equation}
with
\begin{equation}
s_{ab}=-V_{,ab}-\frac{1}{H{\Mpl^2}}\left(V_{,a}\dot\vphi_{b}+\dot\vphi_{a}V_{,b}\right)+\frac{V}{H^2{\Mpl^4}}\dot\vphi_{a}\dot\vphi_{b}.
\end{equation}
And finally, in this gauge the adiabatic metric fluctuations are,
\begin{equation}
\zeta=-\frac{H}{\dot\vphi_{b}^2}\ \dot\vphi_{a}\,\dphi_{a}=-\vn_{a}\dphi_{a}\frac{H}{\sqrt{\dot\vphi_{b}^2}}.\label{zetalin}
\end{equation}
In the slow-roll regime, it can then be written,
\begin{equation}
\zeta=\frac{1}{\Mpl\sqrt{2\epsilon}}\ {\vn_{a}\dphi_{a}},
\end{equation}
where $\epsilon$ is the (first) slow-roll parameter, $2\epsilon\equiv (\Mpl^2 \vn_{a}V_{,a}/V)^2$.
This set of equations provides the necessary ingredient to compute the shape of the power spectrum however
complicated the potential might be. 

To obtain the induced amplitude of the NGs it is however necessary to extend the calculations to higher order
terms. Both the relations (\ref{action2nd}) and (\ref{zetalin}) should then be extended to higher order terms. These calculations were initially carried by Maldacena in Ref. \cite{2003JHEP...05..013M} for a single field. 

The relation between the curvature perturbation and the field fluctuations is in general intricate. The $\delta N$ formalism offers however an insightful way of  representing those couplings (\cite{1982PhLB..117..175S,1996PThPh..95...71S} and see Ref. \cite{2005PhRvL..95l1302L} in the context of nonlinear expansions). Indeed scales of interest being super-Hubble, the curvature can be identified as fluctuation of the number of efoldings from one field trajectory to another. This number can in turn be expanded as a function of the field fluctuations,
\begin{equation}
\delta N(t)=N_{,a}\ \dphi_{a}(t_{*})+\frac{1}{2}N_{,ab}\ \dphi_{a}(t_{*})\dphi_{b}(t_{*})+\dots
\end{equation}
where 
\begin{equation}
N_{,a}=\frac{\partial N(t)}{\partial\dphi_{a}(t_{*})}\ \ \hbox{and}\ \ N_{,ab}=\frac{\partial^2 N(t)}{\partial\dphi_{a}(t_{*})\partial \dphi_{b}(t_{*})}.
\end{equation}
Here $t_{*}$ is a time for which all modes are evolving at super-Hubble scales. This is usually taken as the time of horizon crossing (as in Refs. \cite{2006JCAP...05..019V,2009JCAP...02..017B,2009JCAP...08..016B}) but this is not necessarily so and it can be taken later\footnote{That amounts to compute the nonlinear super-Hubble evolution of the isocurvature fields either separately, from the action expression, or to include it in the $N_{,ab}$ coefficient as illustrated below.}. The presence of a second order term ensures that the resulting metric fluctuations are not Gaussian. This is however not the only possible contribution. Other possibilities come from the fact that the fields themselves may not be Gaussian distributed to start with at time $t_{*}$. The unambiguous way to derive the statistical properties of those fields is to start with the third order action and, using for instance for the ``in-in'' formalism from Schwinger and Keldysh (see Refs. \cite{1961PNAS...47.1075S, Keldysh:1964ud} and \cite{2005PhRvD..72d3514W} for a presentation and use in a cosmological context), derive the properties of the field.

In case of multiple-field inflation the result obtained by Maldacena for single field can be readily extended. The third order action is given by (using the same gauge as before),
\begin{eqnarray}
\mL_{3}&=&
-\frac{1}{6}V_{,abc}\dphi_{a}\dphi_{b}\dphi_{c}\nonumber\\
&&-\frac{1}{4H\Mpl^2}\dot\vphi_{a}\dphi_{a}\left[V_{,bc}\,\dphi_{b}\dphi_{c}+\dot\dphi_{b}\dot\dphi_{b}+\frac{1}{a^2}\partial_{i}\dphi_{b}\partial_{i}\dphi_{b}\right]-\dot\dphi_{a}\partial_{i}\dphi_{a}\partial_{i}\chi\nonumber\\
&&+\frac{1}{8H\Mpl^4}(\dot\vphi_{a}\dphi_{a})^3\left[3-\frac{{\dot\vphi_{a}}^2}{2H^2}\right]+
\frac{(\dot\vphi_{a}\dphi_{a})^2}{4H}\left[\frac{1}{H}\dot\vphi_{b}\dot\dphi_{b}+\partial^2\chi\right]\nonumber
\\
&&-\frac{1}{4H\Mpl^4}\dot\vphi_{a}\dphi_{a}\left[\partial_{i}\partial_{j}\chi\partial_{i}\partial_{j}\chi-\partial^2\chi\partial^2\chi\right],\label{action3}
\end{eqnarray}
where
\begin{equation}
\partial^2\chi=-\frac{V_{,a}}{2H}\dphi_{a}-\frac{1}{2H}\dot\vphi_{a}\dot\dphi_{a}-\frac{V}{2H^2\Mpl^2}\dot\vphi_{a}\dphi_{a}.\label{chiexp}
\end{equation}
In Eqs. (\ref{action3}) and (\ref{chiexp}) the different terms are ordered in terms of the power of the Planck mass. They
can alternatively be ordered with respect to the slow-roll parameters. For single field inflation such a decomposition is unambiguous as described in \cite{2003JHEP...05..013M}. For multiple-field models the decomposition depends on whether the derivatives of the potential have the same order of magnitudes in all field directions. If it is so, the third order action is dominated by the last three term of the second line: they are all of order $\sqrt{\epsilon}H^5/\Mpl$. That corresponds to the expression of the action as derived in \cite{2005JCAP...09..011S}. On the other hand, if only the adiabatic direction has small derivatives then the expansion should be made with more care. {In particular, the first term of the second line
can be of the same order of the other terms of that line; and the very first term can dominate all others inducing \textsl{non-gravity mediated couplings.} Indeed, whereas field fluctuations in the adiabatic direction are intimately coupled to metric fluctuation, and are therefore constrained by slow-roll conditions, this is not the case in transverse directions (provided its effective mass remains below $H$, see discussion in Ref. \cite{2002PhRvD..66j3506B}), and field can develop arbitrarily large couplings. To give another insights into the separation of those terms, one can note that in the de Sitter limit -- which implies that all time derivatives of background quantities vanish -- the only non-vanishing term of $\mL_{3}$ is the first one. It simply describes the self-coupling of fields in an arbitrary expanding background.

\subsection{A geometrical description}

\begin{figure*}
\rightline{ \includegraphics[width=11cm]{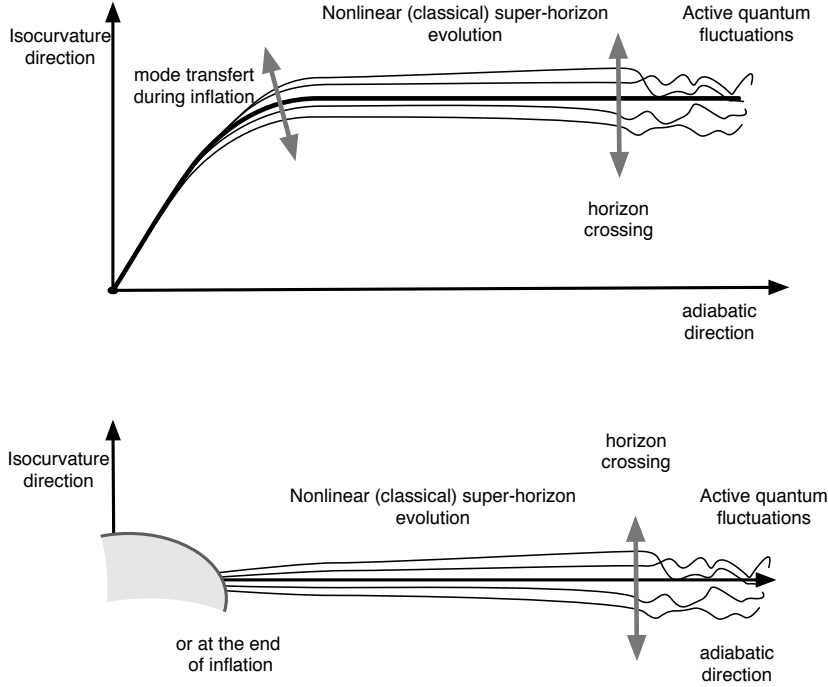}}
\caption{Sketch of the sub- to super-Hubble evolution of the field fluctuations in isocurvature and adiabatic directions and the mechanisms that can lead to significant non-Gaussian metric perturbations. In the top panel, the transfer of modes takes place during the inflationary period while in the bottom at the end of inflation assuming it corresponds to the critical region in grey.}
\label{SuperHTraj}
\end{figure*}

We have now all the necessary ingredients to present the various mechanisms at play in this context as is illustrated in Fig. 
\ref{SuperHTraj}. Those sketches describe field trajectories assuming one isocurvature degree of freedom. The right part of the panels correspond to sub-Hubble dynamics where quantum fluctuations are at play. Non-Gausssianities can be induced at that stage as described by the ``in-in'' formalism. The formula (\ref{SeeryThreePts}) below gives the expression of the field bispectrum but with the assumption mentioned before, that is assuming slow-roll conditions are met in all field directions. It does not therefore encompass all possibilities.

The transfer from isocurvature to adiabatic modes can take place as soon as modes cross the horizon or much later on. In this first case, a full treatment of the metric perturbation with the help of the ``in-in'' formalism is required.
In the latter case further nonlinearities are built during the super-Hubble evolution of the fields. Those evolutions can be taken
into account through the $\delta N$ formalism (which corresponds to a classical evolution of the fields) or in the ``in-in'' formalism applied to later time. Both descriptions are equivalent and should (see Refs \cite{2008PhRvD..78f3534W} and \cite{2009PhRvD..79d3504W} for a general discussion on that) give the same result. 

It is to be noted then that if the super-Hubble evolution of the field is the dominating process, then the bispectrum is going to be local. The later case can be further extended to situations where mode transfer takes place at the very end of inflation. This is the case in particular for extensions of the hybrid models (see Refs. \cite{2003PhRvD..67l1301B,2004PhRvD..70h3004B,2007PhRvD..76d3526B,2008PThPh.120..159S,2009JCAP...02..017B}). Two cases though should be distinguished: when the transfer is itself nonlinear (that would correspond to critical region with a strongly curved boundary on the bottom panel) or the transfer is linear (the boundary is essentially a straight line) in which case the non-Gaussianities are those imprinted in the isocurvature modes.

This latter picture is the one put forward in \cite{2003PhRvD..67l1301B,2007PhRvD..76d3526B} and advocated as a very efficient mechanism for producing non-Gaussianities since it can take advantage of non-gravity mediated mode couplings.

\subsection{Gravity mediated couplings}

When only the leading order in slow-roll parameters is included, and assuming the potential derivatives are of the same order
in all directions, one can derive the field three-point correlation functions (e.g. \cite{2005JCAP...09..011S,2005JCAP...06..003S,2006JCAP...05..019V}),
\begin{eqnarray}
\langle \dphi_{a}({\bk_1}) \dphi_{b}({\bk_2}) \dphi_{c}({\bk_3})\rangle &=&(2\pi)^3\delta_{\rm Dirac}(\vk_{t})\ \mB_{abc}(\vk_{1},\vk_{2},\vk_{3})\nonumber\\
&=& (2\pi)^3 \delta_{\rm Dirac}\left(\vk_{t}\right)
\frac{\Delta_*^2}{\Pi_i k_i^3}  \nonumber\\
&&\hspace{1cm}\times\sum_{\rm perms} \frac{ \mM(k_1,k_2,k_3)}{4 H\Mpl^2}\dot
\vphi_a \delta_{bc} , \label{SeeryThreePts}
\end{eqnarray}
with
\begin{equation} 
\mM(k_1,k_2,k_3) \equiv - k_1 k_2^2 -4 \frac{k_2^2 k_3^2}{k_t} +
\frac{1}{2} k_1^3 + \frac{k_2^2 k_3^2}{k_t^2}(k_2-k_3)~,
\label{mM}
\end{equation}
where $\vk_t=\vk_1+\vk_2+\vk_3$, and it is assumed that all the $k_i$ are of
the same order of magnitude so that they cross the Hubble radius
approximately at the same time. As a result $\Delta^2=k^3 P_{\dphi}(k)$ is, at leading order in slow-roll parameter, the 
same for all three $\vk_{i}$ at horizon crossing. The sum that appears in (\ref{SeeryThreePts}) is over all simultaneous rearrangements of the indices $a$,
$b$, and $c$, and of the momenta $k_1$, $k_2$, and $k_3$ in $\mM $,
such that the relative position of the $k_i$ is respected.
The form presented here for the function $\mM$ is the one given in Ref.
\cite{2006JCAP...05..019V} rewritten from the expression given in Ref. \cite{2005JCAP...09..011S}.

It can be noticed that the resulting field bispectra scale like $1/\Mpl^2$ and are proportional to the field time derivative. As a result, when re-expressed in terms of the slow-roll parameter, they are typically of the order of 
\begin{equation}
\mB_{abc}(\vk_{1},\vk_{2},\vk_{3})\sim\frac{\sqrt{\epsilon}}{\Mpl}\mP_{\dphi}^2\label{SlowRollBispecApprox}
\end{equation}
which implies that the bispectrum of the metric fluctuations is of the order of 
\begin{equation}
\mB_{\zeta}\sim\epsilon\ \mP_{\zeta}^2\label{SlowRollBispecApprox2},
\end{equation}
not taking into account the subsequent growth of mode couplings. What we recover here is qualitatively nothing but the result obtained by Maldacena for single field inflation in Ref. \cite{2003JHEP...05..013M}.

\subsection{Bispectra from non-gravity mediated couplings}

By construction the previous section ignored the non-gravity mediated terms. Here we rather focus on the other limit case, that is we assume that fields in the transverse direction develop large NG through intrinsic couplings (e.g. first term of Eq. (\ref{action3})). So let us consider $N-1$ isocurvature modes $\chi_{I}$ (which is therefore a $N-1$-dimension subspace of the initial $N$-dimension field space) and we keep here the only cubic term of the potential (mainly to be on the same footing with the previous results),
\begin{equation}
\frac{\partial^3 V(\eta,\chi_{1},...,\chi_{N-1})}{\partial \chi_{I}\chi_{J}\chi_{K}}=\frac{\lambda_{IJK}(\eta)}{3!}.
\end{equation}
It is then possible to show explicitly that the $\chi$ fields develop a non-zero bispectra at horizon crossing and after. The amplitude of the coupling is determined by the values of $\lambda_{IJK}$ that a priori have no relation with the slow-roll parameters. It can be shown that it takes the form (see Refs. \cite{2002PhRvD..66j3506B} and \cite{b2010} for the general case),
\begin{eqnarray}
\mB_{IJK}(\vk_{1},\vk_{2},\vk_{3})=-\nu_{IJK}\left[\mP(k_{1})\mP(k_{2})+\hbox{sym.}\right],
\end{eqnarray} 
with, 
\begin{equation}
\nu_{IJK}=\int_{\eta_{*}}^{\eta_{c}}\dd\eta' a^4(\eta')\lambda_{IJK}(\eta')\int_{\eta'}^{\eta}\frac{\dd\eta''}{a^2(\eta'')},
\end{equation}
which, in case of a de Sitter background and when the coefficients $\lambda_{IJK}$ are constant in time, reduces to
\begin{equation}
\nu_{IJK}=\lambda_{IJK}\frac{(N_{c}-N_{*})}{3H^2}.
\end{equation}
This expression is valid for sufficiently late time after horizon crossing. In case of the de Sitter case that simply means that the number of efoldings $N_{c}-N_{*}$ since horizon crossing for the scales of interest is large (i.e.  $N_{c}-N_{*}\gg 1$).
The results that are presented here correspond to cases where the super-Hubble evolution domimates over coupling effects at horizon crossing time. This can be shown explicitly using the ``in-in'' formalism (an early derivation of the three-point correlation function of a test field in a de Sitter background can be found in \cite{1993ApJ...403L...1F}, the four-point correlation in \cite{2004PhRvD..69f3520B}). 

An important consequence of this approximation is that the bispectra take a \textsl{local} form, e.g. the $\vk_{i}$ dependence of the bispectrum is simply that of a product of spectra.

What is then the amplitude of the couplings? In case of a cubic potential one needs the effective mass of the isocurvature mode to be small compared to the Hubble constant. That imposes that the elements of the matrix $1/H \sum_{I} \lambda_{IJK}$ are small compared to unity, the $\dphi_{J}$ being all of the order of $H$\footnote{Another way of arriving to a similar amplitude is to start with a (positive) quartic potential of the shape $\lambda\chi^4$. In this case $\lambda$ (which is then dimensionless) ought simply to be less than unity. Finite volume effect though gives  a non-zero value of $\chi$ corresponding to the one it acquires at the size of our observable universe. Using the Fokker-Planck approach described in \cite{1994PhRvD..50.6357S} it can be shown that it takes typically a value of about $H/\lambda^{1/4}$   \cite{2004PhRvD..70d3533B}. It leads to a non-zero cubic term of similar amplitude.}.

That implies that the bispectra can be up to the order of 
\begin{equation}
\mB_{\chi}(\vk_{1},\vk_{2},\vk_{3})\lesssim\frac{(N_{c}-N_{*})}{H}\left[\mP_{\chi}(k_{1})\mP_{\chi}(k_{2})+\hbox{sym.}\right]
\end{equation}
This is to be compared to Eq. (\ref{SlowRollBispecApprox}). Here, the resulting amplitude is not suppressed by the slow-roll parameter anymore; it is also a factor $\Mpl/H$ larger! That shows how strong non-gravity mediated couplings can be. There is however a missing ingredient: this mechanism can only be efficient \underline{if} a transfer of modes occurs during the inflationary period of the universe, or at the very least at the end.

\section{Multibrid inflation, from toy models to SUSY models}

What the previous analysis has shown is that multiple-field inflation offers plethoras of mechanisms to play with. In the absence of guideline from high-energy physics it seems therefore difficult to extract robust common features. Attempts to overcome this issue 
with the help of high-energy motivated constructions have all, in this context, lead to inflationary models that are hybrid and or extensions of hybrid type inflation. Hybrid inflation (see Ref. \cite{1994PhRvD..49..748L})
is by itself an appealing model; its extensions in the context of multiple-field inflation can lead to interesting phenomenological properties we wish to describe in this section. Let us start with a simple model we will later motivate from super-symmetric (susy) theories. 

\subsection{A toy model}

\begin{figure*}
\rightline{ \includegraphics[width=7cm]{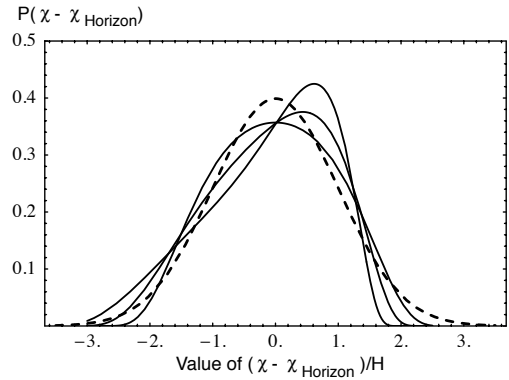}}
\caption{Expected one-point probability distribution distribution of the isocurvature perturbation in case of a quartic potential (solid lines). The plot corresponds to the case $\nu^2(N_{c}-N_{*})/9=1$. Finite volume effects can shift the Hubble size value of $\chi$ to a non-zero value $\chi_{\rm Horizon}$ for our observable universe leading to a skewed distribution of $\chi$. The three solid lines correspond to $\chi_{\rm Horizon}/H=1,\ 0.5$ and $0$. The dashed line is a Gaussian distribution of the same width. See Ref. \cite{2004PhRvD..70d3533B} for details.}
\label{Chi4PDF}
\end{figure*}

Let us consider a simple extension of hybrid inflation with one field (this model has been proposed in Ref. \cite{2003PhRvD..67l1301B}), $\phi$, the inflaton; a light
scalar, $\chi$, that will give birth to isocurvature fluctuations and a third
field, $\sigma$, which is coupled to the two others so that the end of
inflation is triggered when $\sigma$ undergoes a phase transition. To make the model robust in particular with fine tuning in parameters and in the initial conditions, we introduce a quartic coupling in the isocurvature direction so that the potential reads
\begin{eqnarray}\label{threefieldmodel}
 V&=&V(\phi)
        +\frac{\nu^2}{4!}\chi^4+
        \frac{\mu}{2}\left(\sigma^2-\sigma_0^2\right)^2
       +\frac{g}{2}\sigma^2\left(\phi\cos\theta+\chi\sin\theta\right)^2.\label{toymodel}
\end{eqnarray}
Here $\sigma_0$ is the vacuum expectation value (vev) reached by $\sigma$ after the phase transition, $\theta$ is the
mixing angle between $\phi$ and $\chi$ in their coupling to
$\sigma$, $\nu^2$ is the amplitude of the self-coupling. We assume waterfall conditions so that
inflation effectively ends when the effective mass of $\sigma$ vanishes, that
is when
\begin{equation}\label{effectivemass}
  g\left(\phi\cos\theta+\chi\sin\theta\right)^2-2\mu\sigma_0^2=0.
\end{equation}
The value of $\phi$ at the end of inflation is thus
\begin{equation}\label{phiend}
  \phi_{\rm
  end}\equiv\frac{\pm\sqrt{2\mu/g}\,\sigma_0-\chi\sin\theta}{\cos\theta}.
\end{equation}
For $\phi>\phi_{\rm end}$, $\sigma=0$ and the two fields $\phi$ and $\chi$ evolve
independently: $\phi$ drives the inflation while $\chi$ develops non-Gaussianity. The amount of non-Gaussianity of $\chi$ then
depends only on $\nu^2$ and on the total number of e-foldings between horizon crossing and the end of inflation.

When $\theta$ is non-zero, fluctuations of $\chi$ induce metric fluctuations because they change the time at which the phase
transition occurs. Thus the $\chi$-induced metric fluctuations read (assuming $H$ is basically constant during the inflationary period),
\begin{equation}\label{chiinducedR}
  \zeta \simeq H \delta t_{\rm
  end}\simeq-\frac{3H^2}{V_{,\phi}}\,\frac{\sin \theta}{\cos \theta}\,
  \chi.
\end{equation}
Here the transfer of modes is linear and the induced metric fluctuations inherit their non-Gaussian properties from those of the isocurvature modes.

It is to be noted that this model does not require any fine tuning to work, neither in the parameter space (the only requirement is that $\nu$ is less than unity), nor in the initial conditions (if initial conditions are such that $\chi$ is large, this field is bound to roll down near its minimum at a vev of about $H$). It is actually to be noted that, despite the fact that the isocurvature potential is symmetric, finite volume effects induce generically a non zero bispectrum but the amplitude of which depends on the peculiar (patch of the) universe we live in. This mechanism has been described more quantitatively in \cite{2004PhRvD..70d3533B}. The picture we eventually arrive at is that the metric fluctuation is the superposition of a Gaussian field and a non-Gaussian field the latter being obtained essentially from a local nonlinear transformation\footnote{Note that strictly speaking it can be shown that the tree order correlations of the induced $\chi$ field are identical by those induce by such a transform (see Refs. \cite{2002PhRvD..66j3506B,b2010}). If radiative corrections were to be taken into account the above statement would not be correct.} of another Gaussian field of the same variance,
\begin{equation}\label{chiinducedR2}
  \zeta \simeq-\left.\frac{3H^2}{V_{,\phi}}\right\vert_{t=t_{*}}\,
  \dphi-\left.\frac{3H^2}{V_{,\phi}}\right\vert_{t=t_c}\,{\tan \theta}\,\mG[\chi_{0}],\end{equation}
  with
\begin{equation}
\mG[\chi_{0}]=\chi_{0}\left[1+\frac{(N_{c}-N_{*})\nu^2}{9H^2}\chi_{0}^2\right]^{-1/2}
\end{equation}
where $\chi_{0}$ is a Gaussian field of variance $H_{*}$ (but not necessarily of zero mean).
Fig. \ref{Chi4PDF} illustrates the expected shapes of the one-point probability distribution function (PDF) of $\chi$ that one obtains from such a transform.

\subsection{Protecting masses in multiple-field inflation}

From a high-energy physics point of view, the introduction of such multiple degrees of freedom raises a number of questions. Which constructions are more natural? Are such models robust in a high-energy physics context? In particular the protection of the masses and coupling constants of the scalar degrees of freedom we introduced is critical for the viability of the models. If large-scale divergences can be taken care of, mainly from horizon effects or invoking the semi-classical approach described by the Fokker-Planck equation,\cite{1994PhRvD..50.6357S},  masses and coupling constant are a priori unprotected from small-scale divergences, a problem encountered  in expanding backgrounds as well as in Minkowski space-time.  In case of a quartic potential, it is always possible to invoke renormalization properties but it then requires fine-tuning. To avoid it it has been shown in Ref. \cite{2005PhRvD..71f3529B} that if scalar degrees of freedom were imbedded in a super-symmetric multiplet, their power spectrum - and as a consequence their masses - are protected from radiative corrections. 
Construction in a susy context is then a natural playground for the derivation of viable models and one reason that makes models such as $D$-term or $F$-term inflation described respectively in Ref. \cite{1996PhLB..388..241B} and in Ref. \cite{1994PhRvL..73.1886D} attractive.

\subsection{A viable model in the context of D-term super-symmetry}

It goes beyond the scope of this short review to describe the mechanisms at play in the construction of the $D-$term inflation (see for instance Ref. \cite{2000cils.book.....L}). It is based of the up-lifting of flat direction due to (weak) radiative corrections when the current vacuum breaks a supersymmetric Lagrangian. In case of $D$-term infation, the whole structure derives from the expression of the superpotential that in this case takes the form
\begin{equation}
W=\lambda\,\mS\phib\phi
\end{equation}
where $\mS$, $\phib$ and $\phi$ are neutral or charged fields and $\lambda$ a dimensionless coupling constant. 
The resulting potential is
\begin{eqnarray}
V&=&V_{\rm{1-loop}}(\mS)
+\lambda^2\left\vert \mS\right\vert^2\vert\phib\vert^2+
\frac{g^2}{2}\left(-\vert\phib\vert^2+\xi\right)^2
\end{eqnarray}
where $g$ is a gauge field coupling constant, $\xi^{1/2}$ if the energy scale provided by the Fayet-Iliopoulos term and which depends on the (complex scalar) fields $\mS$ and $\phib$.  Here $V_{\rm{1-loop}}$ is the contribution due to
the radiative corrections (computed from the formula of Coleman and Weinberg, \cite{1973PhRvD...7.1888C}) that ensures a slow rolling of $\mS$ towards its minimum. One recognizes here a hybrid model. 

But then, following \cite{2007PhRvD..76d3526B}, nothing prevents the introduction of  multiple light fields $\mS_i$ coupled to
the same charged $U(1)$ fields,
\begin{equation}
W=\sum_{i}\frac{\nu_{i}}{3}\mS_{i}^3+\lambda\left(\sum_{i}\alpha_{i}\mS_{i}\right)\phib\phi
\end{equation}
where $\nu_{i}$ and $\alpha_{i}$ are dimensionless parameters.
Obviously if a peculiar self-coupling parameter $\nu_i$ is small enough, the corresponding $\mS_i$ field can participate in the 
inflaton (depending on its initial vev). The corresponding upper bound for $\nu_i$  for such a possibility to occur comes from the fact that when the vev of $\mS_i$ is below the Planck scale, 
the contribution of the quartic potential it induces should be
negligible against the radiative correction terms. It leads to the
constraint,
\begin{equation}
\nu^2_{i}\ll \lambda^4.
\end{equation}
The fields for which $\nu_{i}$ is above this bound will rapidly roll towards the origin
but they still can  develop  significant super-Hubble fluctuations as long as $\nu_{i}$
is smaller than unity.

The resulting expression of the potential corresponds to the following effective form,
\begin{eqnarray}
V&=&V_{\rm{1-loop}}+\nu_2^2\vert\mS_2\vert^4
+\lambda^2\left\vert \cos\theta\mS_{1}+\sin\theta\mS_{2}\right\vert^2\vert\phib\vert^2
\nonumber\\
&&+
\frac{g^2}{2}\left(-\vert\phib\vert^2+\xi\right)^2
\end{eqnarray}
involving the (complex scalar) fields $\mS_1$, $\mS_2$ and $\phib$. It corresponds to the model (\ref{toymodel}) with $\vphi\equiv \mS_{1}$, $\chi=\mS_{2}$ with the 
difference that the fields are complex (and thus carry two degrees of freedom each). It changes the resulting phenomenology
only in details (for instance on the expected amplitude of the trispectrum, see Ref. \cite{2007PhRvD..76d3526B}).


\begin{figure*}
\rightline{ \includegraphics[width=5.2cm]{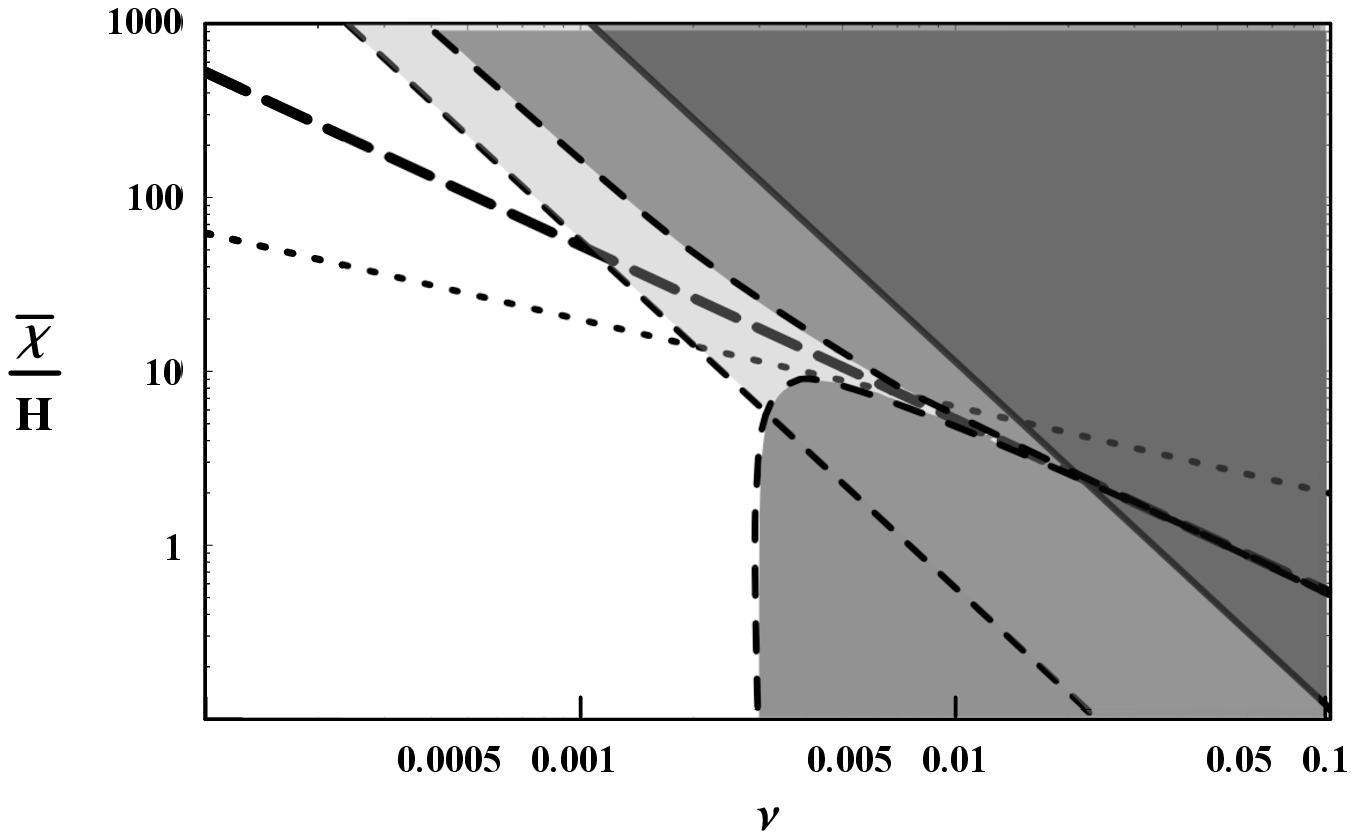} \includegraphics[width=5.2cm]{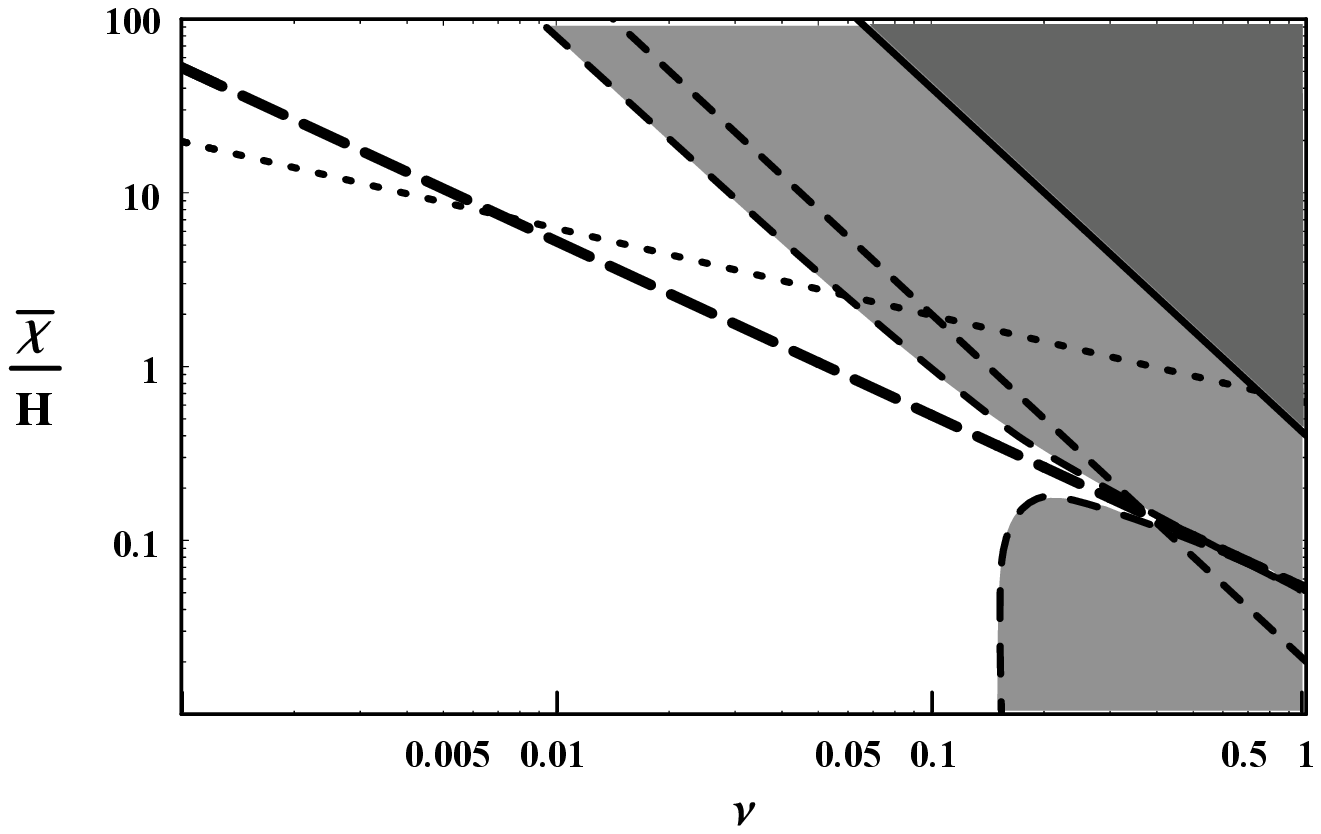}}
\caption{Exclusion diagrams for parameters $\nu$ and $\chib$ for $\theta=\pi/4$ (left panel) and for $\theta=0.1$ (right panel). The gray areas are the exclusion zones, dark gray corresponding to the constraints derived from WMAP and medium and light gray for the constraints expected from Planck. The straight line boundaries correspond to bispectrum observations  (of slope $-2$), the other case to the trispectrum. The long dashed is the location where the terms contributing to the trispectrum cancel. We adopted the results of \cite{2006PhRvD..73h3007K} on the upperbounds the Planck mission is expected to provide, $f_{\rm NL}=5$ and $\tau_{\rm NL}=560$. The dotted lines are the locations where $\chib$ is equal to its expected 1-$\sigma$ fluctuation. }
\label{ExclusionMap}
\end{figure*}

The entire non-Gaussian properties of such a model are determined by three parameters that are,
\begin{itemize}
\item \underline{the coupling constant $\nu$} that sets the amplitude of the non-Gaussianities that can be reached in the isocurvature mode. For perturbation theory to be valid, $\nu^2$ should be less than unity. Actually we found that the combination $N_{e}\nu^2$, where $N_{e}$ is the number of efoldings since horizon crossing, ought to be small. This condition ensures that the field $\chi$ is light compared to the Hubble constant, its effective mass is $\nu^2 H^2$, and that its evolution is perturbative all through its super-Hubble evolution. 
\item \underline{the mixing angle $\theta$} that determines the fraction of isocurvature modes that eventually enters the metric perturbations. This angle can be anywhere between $0$ and $\pi/2$. 
\item \underline{the super-Hubble value of $\chi$}, $\bar\chi$, is an priori undetermined value which depends on the patch of the universe we live in. Within some hypothesis, its cosmic PDF can be determined and therefore its expected values. Such derivation is done with the help of the Fokker-Planck equation (see Ref. \cite{2007PhRvD..76d3526B} in this multi-dimension context). The value $\bar\chi$ determines in particular the amplitude of the expected bispectrum (it is zero if $\bar\chi$ is set to 0) and the relative contribution of the terms contributing to the trispectrum.
\end{itemize}

On Fig. \ref{ExclusionMap}, one can see the current and expected constraints on parameter space that can be put on such a model in the $\nu-\chib$ plane for two different values of $\theta$. The upper right parts of the diagram are a priori excluded. It can be seen that bispectrum and trispectrum play a complementary role in constraining these models.

\section{Conclusions}

The exploration of the various types of coupling terms that appear in the action leads to distinguish gravity from non-gravity mediated couplings. In one-field inflation, because the field fluctuations and the metric fluctuation are locked together, only the former
can be found. In multiple-field inflation however this is not necessarily so and it opens the possibility of having a richer phenomenology. Whereas the gravity mediated couplings are ubiquitous but induce only modest effects, the non-gravity mediated couplings can be very efficient, although nothing ensures that they are generically at play.
Giving the freedom one has in building potentials it is however certainly possible to design models exhibiting any kind of scale and geometrical dependences in the bispectrum (this has been tentatively explored in Ref. \cite{2009arXiv0911.2780B}), but actual constructions motivated by high-energy physics
point to hybrid inflation type models where couplings are eventually of local type. The extended $D$-term susy model presented here is an example of such a construction. A few lessons can be drawn from its analysis, 
\begin{itemize}
\item in such a susy framework masses and coupling constants are naturally protected. This is at the expense of the doubling of the number of scalar degrees of freedom; 
\item this extension of $D$-term inflation leads to hybrid models where the end line, the critical line where the inflationary period terminates, is linear in all field directions. This is at variance with the construction proposed in \cite{2008PThPh.120..159S} where NGs originate from a nonlinear critical line;
\item in such a model there is no need for specific fine tunings, neither in the initial conditions nor in the parameter space. It is also to be noted that such models induce a dumping of the rare event tails - on both sides.
\item the transfer of modes, from isocurvature to adiabatic direction, takes place at the time the inflation stops. It leads to local types for bispectrum, and trispectrum, shapes.
\item calculations were carried here at leading order in perturbation theory. The conditions for such calculations to be valid during the sub-Hubble evolution is that the coupling constant $\nu^2$ is less than unity. Once the evolution is super-Hubble however the amplitude of the fluctuation couplings is driven by $\nu^2 (N-N_{*})$ which ought then to be small. There is therefore a regime where perturbation theory can be used during sub-Hubble evolution but not during the whole super-Hubble evolution. That could lead to a new phenomenology; e.g. to nonlocal effects in the shape of the high order spectra.
\end{itemize}
The model at hand is therefore rather sound with not so much freedom in its predictions. But this is by no means exhaustive. Other scenarios are certainly possible.

\section*{Acknowledgments} The author is grateful to J-Ph. Uzan and F. Vernizzi for a careful reading of the manuscript and discussions during the writing of this review and to M. Sasaki and D. Wands for their invitation to write it. 

\section*{References}

\bibliography{EarlyUniverse}

\end{document}